# Growth optimization and device integration of narrow-bandgap graphene nanoribbons

Gabriela Borin Barin[1]*, Qiang Sun[1]*+, Marco Di Giovannantonio[1++], Cheng-Zhuo Du[2], Xiao-Ye Wang[2], Juan Pablo Llinas[3], Zafer Mutlu[3], Yuxuan Lin[3], Jan Wilhelm[4], Jan Overbeck[1], Colin Daniels[5], Michael Lamparski[5], Hafeesudeen Sahabudeen[6], Mickael L. Perrin[1], José I. Urgel[1+++], Shantanu Mishra[1++++], Amogh Kinikar[1], Roland Widmer[1], Samuel Stolz[1+++++], Max Bommert[1], Carlo Pignedoli[1], Xinliang Feng[6], Michel Calame[1], Klaus Müllen[7,8], Akimitsu Narita[7,9], Vincent Meunier[5], Jeffrey Bokor[3], Roman Fasel[1,10] and Pascal Ruffieux[1]

[1]Empa, Swiss Federal Laboratories for Materials Science and Technology, 8600 Dübendorf, Switzerland

[2]State Key Laboratory of Elemento-Organic Chemistry, College of Chemistry, Nankai University, Tianjin 300071, China

[3]Department of Electrical Engineering and Computer Sciences, University of California, Berkeley, CA 94720, United States

[4] Institute of Theoretical Physics, University of Regensburg, D-93053 Regensburg, Germany

[5]Department of Physics, Applied Physics, and Astronomy, Rensselaer Polytechnic Institute, Troy, New York 12180, United States

[6]Center for Advancing Electronics Dresden, Department of Chemistry and Food Chemistry, TU Dresden, Dresden 01062, Germany

[7]Max Planck Institute for Polymer Research, 55128 Mainz, Germany

[8]Department of Chemistry, Johannes Gutenberg-Universität Mainz, 55128 Mainz, Germany

[9]Organic and Carbon Nanomaterials Unit, Okinawa Institute of Science and Technology Graduate University, 1919-1 Tancha, Onna-son, Okinawa 904-0495, Japan

[10]Department of Chemistry, Biochemistry and Pharmaceutical Sciences, University of Bern, 3012 Bern, Switzerland

+Present address: Materials Genome Institute, Shanghai University, Shanghai 200444, China

++Present address: Istituto di Struttura della Materia – CNR (ISM-CNR), via Fosso del Cavaliere 100, Roma 00133, Italy

+++Present address: IMDEA Nanoscience, C/Faraday 9, Campus de Cantoblanco, Madrid 28049, Spain

++++Present address: IBM Research – Zurich, Rüschlikon 8803, Switzerland

+++++Present address: Department of Physics, University of California, Berkeley, Berkeley, CA 94720, USA

*These authors contributed equally to this work

**Abstract**

The electronic, optical and magnetic properties of graphene nanoribbons (GNRs) can be engineered by controlling their edge structure and width with atomic precision through bottom-up fabrication based on molecular precursors. This approach offers a unique platform for all-carbon electronic devices but requires careful optimization of the growth conditions to match structural requirements for successful device integration, with GNR length being the most critical parameter. In this work, we study the growth, characterization, and device integration of 5-atom wide armchair GNRs (5-AGNRs), which are expected to have an optimal band gap as active material in switching devices. 5-AGNRs are obtained via on-surface synthesis under ultrahigh vacuum conditions from Br- and I-substituted precursors. We show that the use of I-substituted precursors and the optimization of the initial precursor coverage quintupled the average 5-AGNR length. This significant length increase allowed us to integrate 5-AGNRs into devices and to realize the first field-effect transistor based on narrow bandgap AGNRs that shows switching behavior at room temperature. Our study highlights that optimized growth protocols can successfully bridge between the sub-nanometer scale, where atomic precision is needed to control the electronic properties, and the scale of tens of nanometers relevant for successful device integration of GNRs.

**Introduction**

Graphene nanoribbons (GNRs) are strips of graphene with a distinct number of carbon atoms in width, with exciting properties deriving from quantum confinement and related bandgap tunability[1]. The ability to modify GNRs' electronic and magnetic properties by structural design at the atomic level makes them an ideal platform for numerous device applications ranging from classical transistors to spintronics and photonics[2–5].

Atomically precise GNRs have been successfully synthesized by surface-assisted polymerization and cyclodehydrogenation of specifically designed precursor molecules. Since the pioneering work of Cai et al in 2010[2], GNRs with armchair edges and different widths[6–10], with zigzag[11] and cove[12] edges, as well as edge-extended GNRs hosting topological quantum phases[13,14] have been reported.

For the realization of devices exploiting their exciting electronic and magnetic properties, GNRs need to be transferred from the growth substrate, usually noble metals, onto technological relevant ones such as silicon with various types of oxides[15,16]. Because of their robustness and stability under ambient conditions, AGNRs have been the most studied class of nanoribbons. AGNRs can be classified into three families according to the number of carbon atoms $N$ across the GNR width $N = 3p$, $3p + 1$ and $3p + 2$, where $p$ is an integer[1]. The family of $3p + 2$ possesses the lowest band gap[17,18], and is therefore most promising for high-performance electronic switching devices. However, to date, only wide-bandgap GNRs, such as 7-, 9-, 13-AGNRs revealed switching behavior when integrated in field-effect transistor (FET) devices[19–21]. Narrow-bandgap nanoribbons available so far had limited length, increased chemical reactivity or showed weak coupling to contacts, and consequently showed high contact resistances, all of which has limited their successful application in devices[17,21–23].

The first member of the $3p+2$ family, 5-AGNR, has first been synthesized and studied by Kimouche et al[6], who observed a very small electronic gap of around 100 meV on Au(111). In a later study, Lawrence et al[18] investigated the length-dependent electronic structure of 5-AGNRs and unveiled slowly decaying topological end states localized at the ribbon termini to be responsible for the seemingly low band gap reported in the previous study. The end states in finite-length 5-AGNRs were also discussed by El Abbassi et al[17], when investigating their quantum dot behavior in a FET-device configuration. Due to the slow decay of these in-gap states into the bulk of the GNRs, FET-devices bridged by short 5-AGNRs as the active material showed a metallic behavior at room temperature and addition energies in the range of 100-300 meV at 13 K.

The length of 5-AGNR samples produced so far has been limited to a few nanometers only, which imposes a particular challenge for device fabrication. To date, only graphene electrodes with sub-5 nm nanogaps created by electrical breakdown have successfully bridged the short 5-AGNRs[17]. However, the fabricated FET devices did not show switching behavior at room temperature[17]. Growth conditions yielding longer 5-AGNRs would not only be beneficial for their integration into FET devices with traditional metal electrodes, whose minimum separation is about 20 nm[21], but would also allow greater overlap between GNRs and contacts to reduce contact resistance.

Here, we report new growth protocols to extend the maximum length of 5-AGNRs from below 10 nm to 45 nm and their successful integration into devices using metal electrodes with source-drain gaps of 15 – 20 nm. The improved growth protocols build on the use of Iodine-substituted precursors and optimized precursor coverages. Using scanning tunneling microscopy (STM) and temperature-programmed X-ray photoelectron spectroscopy (TP-XPS), we systematically investigated the impact of the precursor halogen functionalization (Br versus I) and initial coverage to unveil the growth mechanism of 5-AGNRs. Moreover, we demonstrate the robustness of 5-AGNRs by transferring them to different substrates and exploit Raman and UV-vis spec-troscopies to identify signatures of ribbons with different lengths. Finally, 5-AGNR-FET devices with metal electrodes were fabricated and showed switching behavior at room temperature, confirming their semiconductor character.

**Results and Discussion**

**STM and TP-XPS**

STM and TP-XPS are two highly complementary techniques. While the former provides real-space information with sub-nanometer resolution, insight into the reaction kinetics with high temperature resolution can be obtained from the latter[24,25]. Here, we applied the two techniques to clarify the influence of precursor coverage and the type of halogen substitution on the length distribution of 5-AGNRs.

To obtain 5-AGNRs samples, the precursor monomers are deposited on a Au(111) substrate, followed by an annealing step to activate dehalogenative aryl-aryl coupling and subsequent cyclodehydrogenation. Two samples with different molecular coverages (~ 0.9 and ~1.5 monolayer, ML, hereafter called low- and high-coverage, respectively) are prepared for both dibromo-perylene (DBP)[6] and diiodoperylene (DIP) precursors (Figure 1a) (for details on the synthesis of DIP see Methods and SI, S1-S3). Representative STM images of 5-AGNRs grown from low and high coverage DBP (DIP) are shown in Figure 1b,d (Figure 1c,e), respectively. The average length of 5-AGNRs for the low-coverage cases is ~3 nm. The length-distributions are largely dominated by dimers and trimers, as revealed by the histograms in Figure 1f,g (red traces) and are comparable to those reported previously[6,18]. On the other hand, the average length of 5-AGNRs in the high-coverage regime (Figure 1d,e) is significantly increased. It reveals a striking difference in average length for 5-AGNRs grown from DBP and DIP, which are ~5 and ~17 nm, respectively (blue traces Figure 1f,g). A closer inspection of the STM images of the high coverage samples indicates the presence of a few second layer 5-AGNRs formed from the DIP (Figure 1e, see blue arrow), in contrast to the case of DBP, where no 5-AGNRs on top of the first layer are observed (Figure 1d).

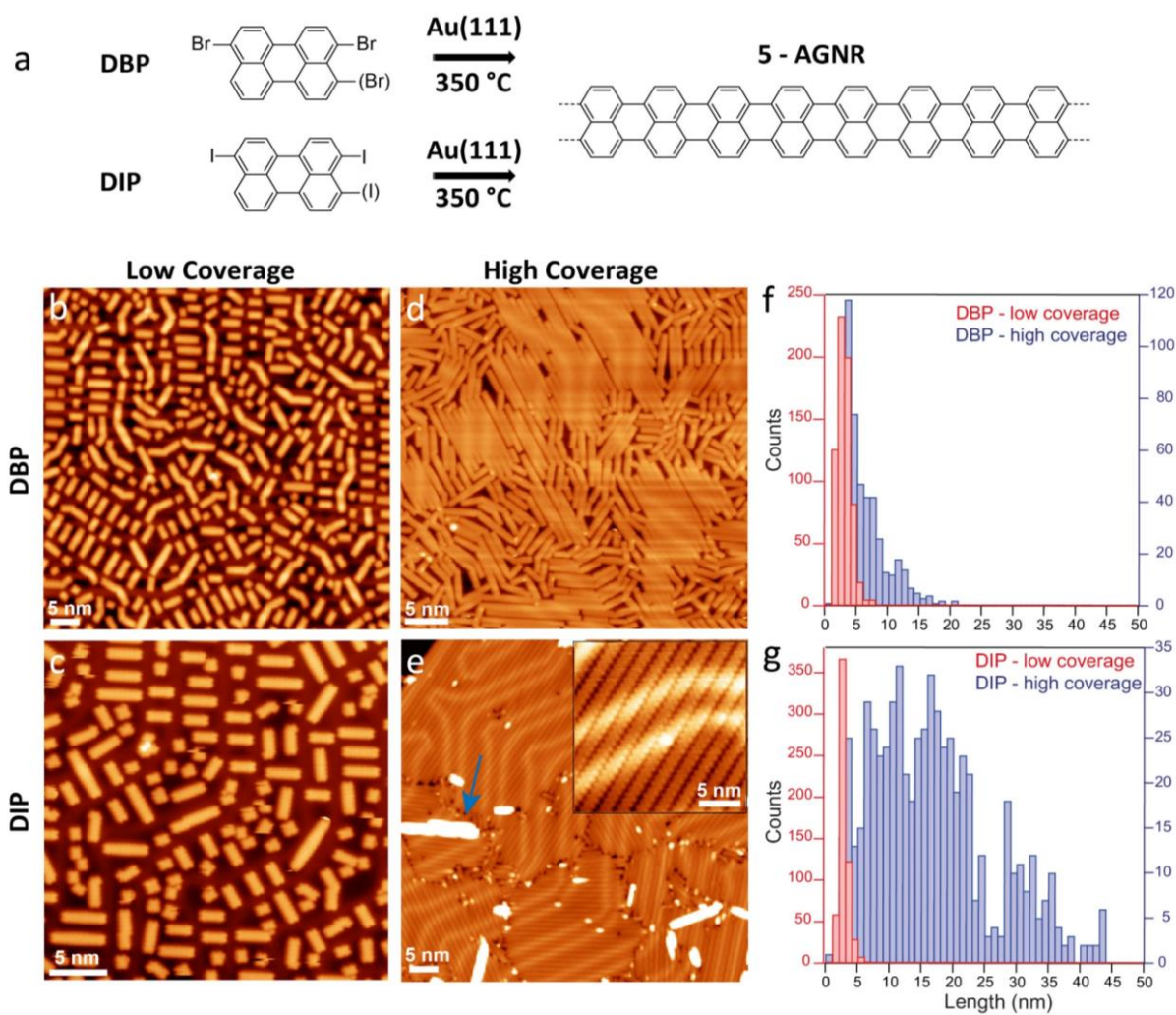

**Figure 1.** On-surface synthesis of 5-AGNRs from two different precursors. (a) On-surface synthetic route to the formation of 5-AGNRs from the dibromoperylene (DBP) and diiodoperylene (DIP) precursors. Overview STM images of low coverage samples (0.9 ML) synthesized from (b) DBP ($V_s = 1$ V, $I_t = 60$ pA) and (c) DIP ($V_s = -1$ V, $I_t = 80$ pA). Overview STM images of high coverage samples synthesized from (d) DBP ($V_s = -1$ V, $I_t = 50$ pA), and (e) DIP ($V_s = -1.5$ V, $I_t = 200$ pA), inset: $V_s = -0.05$ V, $I_t = 0.2$ nA. The blue arrow indicates 5-AGNRs on top of the first layer. (f,g) Histograms showing the length distributions of 5-AGNRs synthesized from DBP and DIP starting from samples with different molecular coverage, respectively.

Although high-resolution STM imaging provides detailed structural information on the prepared samples[6,18], this technique is limited with respect to real-time monitoring of thermally activated reactions. To identify the intermediates and unravel growth mechanisms as well as formation kinetics of 5-AGNRs, we carried out TP-XPS measurements, i.e., repeatedly recorded XPS spectra of the relevant core level signals during the annealing of the sample from -50 to 450 °C (heating rate of 0.2 °C·s$^{-1}$) (Figure 2). This investigation tracks on-surface reactions and unambiguously identifies the sequence of relevant reaction steps[24,26–31]. The molecu-lar precursors were deposited onto a cold substrate (below -120 °C) to prevent their deiodination[32]. We monitored C 1s, and either Br 3d for DBP or I 4d for DIP core level signals of four samples, one with high and one with low initial precursor coverage for both, DBP and DIP. Each horizontal line of the TP-XPS maps shown in Figures 2a-d represents a single XPS spectrum of the examined core level.

As displayed in the Br 3d and I 4d maps in Figure 2, distinct chemical shifts of the Br 3d and I 4d doublets towards lower binding energy (BE) are discernible and attributed to the dehalogenation of the molecular precursors and chemisorption of the halogens on the gold substrate, forming gold-halides (i.e., from C-X to Au-X, with X representing either Br or I)[24]. By extracting intensity *vs.* temperature curves of the Au-X signals from the TP-XPS maps (shown in Figure 2e and 2f for DIP and in Figure S4 for DBP), we can identify the onset temperatures of deiodination (low coverage -20 °C; high coverage 0 °C) and debromination (low coverage +70 °C; high coverage +90 °C) from the decrease of the C-X and simultaneous increase of the Au-X signal. Further, the reduction of the Au-X signal at higher temperatures corresponds to halogen desorption, which is completed at 380 °C (350 °C) and 320°C (380 °C) for low and high coverage DIP (DBP), respectively (Figure 2e and 2f for DIP low and high coverage, respectively and S4 for DBP).

From the TP-XPS maps of the halogen core levels, we can conclude that high initial molecular coverage slightly increases the onset temperature of dehalogenation, which we tentatively ascribe to: (i) a limited molecule mobility and (ii) a lower ratio of molecule per catalytically active site. Moreover, deiodination occurs at considerably lower temperatures compared to debromination due to the lower bond dissociation enthalpy for C-I with respect to C-Br[33,34]. This strongly impacts the second molecular layer, causing molecular desorption before dehalogenation for DBP, as evidenced in the significant intensity drop at 90 °C of the overall Br 3d signal in Figure 2b.

To gain more insight into the reaction mechanism, we now focus on TP-XPS maps of the C 1s core level in Figure 2. Chemical shifts towards lower BE are observed at temperatures coinciding with the previously identified dehalogenation process. From the high-resolution (HR) XPS spectrum of the low coverage DIP sample after completed dehalogenation (115 °C, Figure S5b), we identify the formation of an organometallic (OM) compound, as recently observed by Berdonces-Layunta *et at*[35]. Upon further annealing, the C 1s spectra shift to higher BE (Figure 2c, HR-XPS in Figure S5c), which implies the transformation of OM structures into 5-AGNR. The OM phase is stable over a wide temperature range when using DIP as a precursor (from 70 to 150 °C). This allowed us to acquire STM images of this OM phase (i.e., after annealing to 115 ˚C), which clearly revealed the presence of OM chains on the surface, spaced by chemisorbed iodine atoms (Figure S6e,f). The formation of C-Au-C organometallic chains is also present but less obvious for the DBP precursor (STM image in Figure S6c). There is only a small shift toward lower BE visible in the TP-XPS map, which suggests a significant overlap between the debromination and cyclodehydrogenation processes. Notably, we did not observe the C-C linked poly-perylene before forming 5-AGNR from neither XPS spectroscopy nor STM imaging for both DBP and DIP precursors, similarly to the case of the debrominative homo-coupling of tetrabromonaphthalene on Au(111)[36].

Concerning the temperature evolution of the C-Au-C intermediate, we focused on the C 1s TP-XPS maps of DIP in Figure 2c,d, because the reduced temperature difference between debromination and cyclodehydrogenation in the DBP case prevented a reasonable fit to be performed. For the low coverage case, Figure 2e, the intensities of the C-Au-C component decrease (i.e., cyclodehydrogenation takes place) above 130 °C, while for the high coverage sample this happens above 160 °C, Figure 2f. Since the transformation of the OM chain into 5-AGNRs is accompanied by a cyclodehydrogenation process, the detached iodine atoms may combine with the released hydrogen atoms and desorb as HI, as previously shown by temperature-programmed desorption measurements in 9-AGNRs[25]. Remarkably, apart from the different onset temperatures of the cyclodehydrogenation for the DIP case at low (130 °C) and high (160 °C) coverages, the slopes of the kinetic curves are different in the way that for the low coverage sample, the weight of the organometallic phase decays more rapidly (130 to 200 ˚C) than for the high coverage sample (160 to 320 ˚C), as indicated by the green areas in Figures 2e and 2f, respectively. For the high coverage case, the temperature range over which the cyclodehydrogenation process takes place coincides with the halogen desorption.

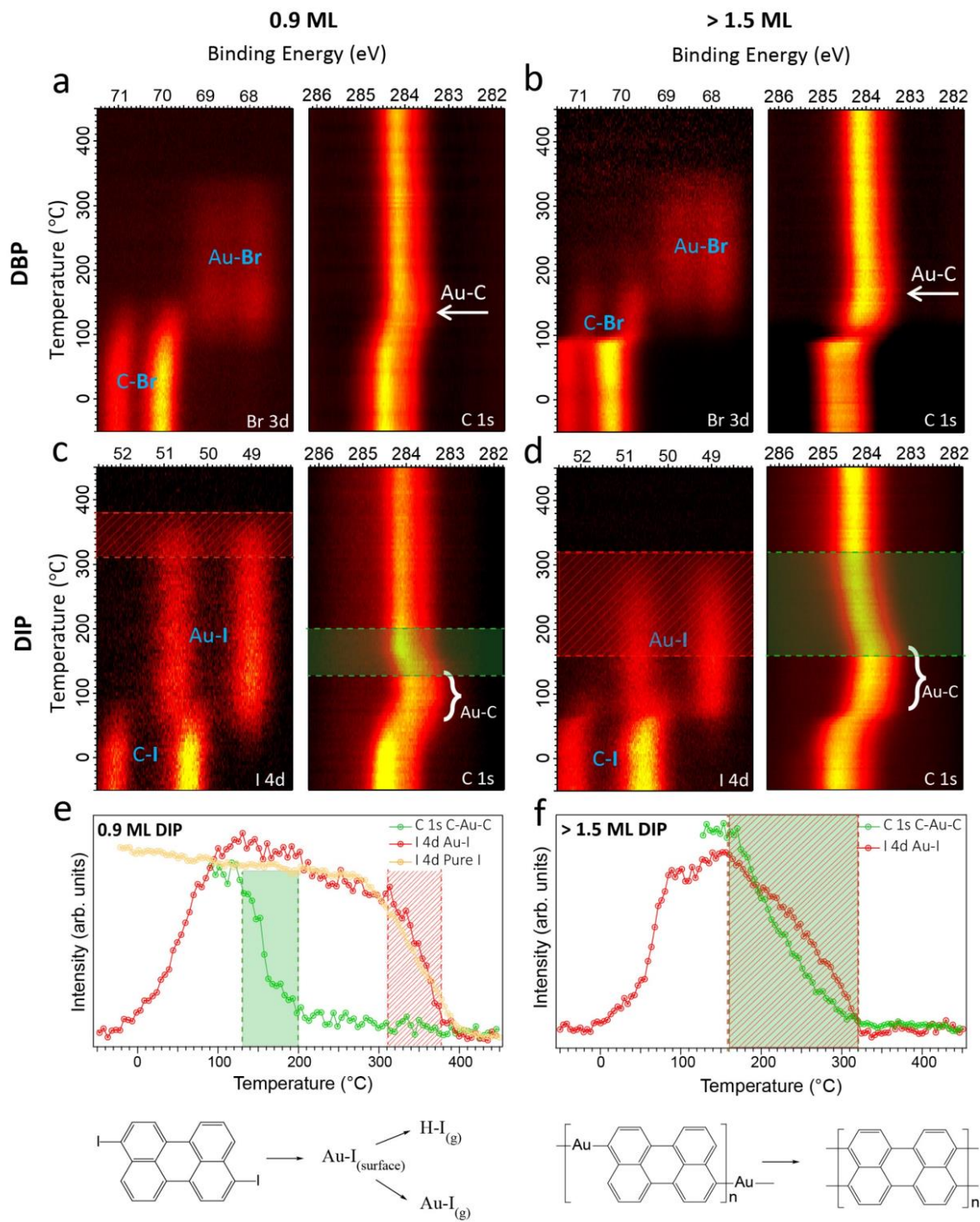

**Figure 2**. TP-XPS maps and kinetic curves describing the on-surface reactions. TP-XPS maps of Br 3d and C 1s core levels recorded during the heating (rate of 0.2 °C·s$^{-1}$) of the (a) low (~0.9 ML) and (b) high (>1.5ML) coverage DBP samples. (c,d) TP-XPS maps of I 4d and C 1s core levels recorded during the heating (rate of 0.2 °C·s$^{-1}$) of the (c) low (~0.9 ML) and (d) high (>1.5ML) coverage DIP samples. The maps with high precursor coverages in (b) and (d) are normalized by the total area to highlight the energy shift. (e) DIP low coverage: Intensities of the Au-I component extracted from the I 4d signal and the intensities of the OM intermediates obtained from fits of the individual spectra of the C 1s TP-XPS maps in (c). The kinetic curve of pure iodine on Au(111) is included for comparison (yellow curve, from ref 25). (f) DIP high coverage: Intensities of the Au-I signals extracted from the I 4d and the intensities of the OM intermediates obtained from fits of the individual spectra of the C 1s TP-XPS maps in (d). Green and red areas in (c), (d), (e), and (f) highlight the relevant transitions discussed in the text. The reaction schemes shown at the bottom indicate different states of the iodine and molecules during annealing.

Based on the experimental observations reported above, we provide a tentative description of the microscopic processes to explain the strongly increased 5-AGNR length obtained for the DIP at high coverage. The similar range for the iodine desorption and cyclodehydrogenation for the DIP at high coverage (red and green areas, respectively in Figures 2d and 2f) suggests that the two processes are closely connected. In fact, hydrogen atoms released during cyclodehydrogenation most likely combine with the detached iodine atoms present on the Au(111) surface, promoting the associative desorption as HI[37,38]. Therefore, iodine atoms can be seen as traps for atomic hydrogen, which reduces the likelihood of hydrogen-based radical passivation during GNR growth.

The efficiency of this trapping process is increased by the presumed proximity of iodine and hydrogen due to the densely packed molecular layer as well as a continuous hydrogen supply. The presence of a molecular reservoir at the second layer provides additional species to fill the empty spaces that appear due to the shrinking of the OM phase into GNRs, maintaining a high molecular density in the first layer. On the other hand, at low coverage, GNRs are formed in a narrow temperature/time range (green area in Figure 2e), releasing abundant hydrogen atoms that can easily diffuse on the surface (due to the lower molecular density) and passivate the radicals of growing ribbons. The excess hydrogen desorbs (most likely as $H_2$) without affecting the iodine atoms that remain on the Au(111) surface. The iodine desorption kinetics from low coverage DIP samples resembles the one of pure iodine (red and yellow curves in Figure 2e, respectively), suggesting iodine desorption as AuI as proposed in a previous work[25].

**Raman Spectroscopy**

Raman spectroscopy has been widely used to characterize carbon nanomaterials due to its ease-of-use, high throughput, and sensitivity to structural details.[39,40] GNRs present well-known Raman fingerprints that allow evaluation of their structural quality and length before and after substrate transfer as well as after device processing[15,17,21,40,41]. The Raman spectrum of GNRs is dominated by three main sets of peaks: the so-called G, CH/D, and radial breathing-like (RBLM) modes. As previously observed for graphene and carbon nanotubes, the G mode originates from in-plane vibrations of the $sp^2$ lattice[42] and is usually found around 1600 cm$^{-1}$. The CH/D modes are unique features due to the presence of (hydrogen-passivated) edges in GNRs that break the periodicity of the perfect $sp^2$ honeycomb lattice and are typically present in the spectral range of 1100-1400 cm$^{-1}$. Lastly, the RBLM is unique to GNRs and it originates from the transverse acoustic phonon and its frequency sensitively depends on the width of armchair GNRs[15,43]. Very recently, we showed that some features of the Raman spectra are also sensitive to the length of GNRs[40]. By investigating ribbons of different lengths, we detected a low-energy peak, the frequency of which is solely dependent on the GNR length, the so-called longitudinal compressive mode (LCM).

Here, the growth protocol-dependent average length of 5-AGNRs is used to investigate the length-dependence of the LCM fingerprint of 5-AGNRs (Figure 3a). While for long 5-AGNRs with an average length of 20 perylene units (obtained on the high coverage sample) the LCM was observed at ~100 cm$^{-1}$, short 5-AGNRs, mainly perylene dimers and trimers (obtained on the low coverage sample), revealed LCM contributions at 124 and 187 cm$^{-1}$.

Next, we focus our analysis on the high-frequency range of the Raman spectrum in which we observe clear differences for the G peak for short and long 5-AGNRs. The G peak of GNRs is composed of two in-plane optical modes: a transverse-optical (TO) and a longitudinal-optical (LO) mode[44]. The LO and TO modes are present in all AGNR families and were previously resolved experimentally for both 7- and 9-AGNRs, with their relative intensities being strongly dependent on the excitation energy[15].

In Figure 3a we show the Raman profiles of long and short 5-AGNRs, both measured with a wavelength of 785 nm (1.58 eV, see also Figure S7 for UV-Vis characterization from 200-800 nm). The experimental Raman profile for short 5-AGNRs (red) reveals two peaks in the G region, 1531 cm$^{-1}$ and 1563 cm$^{-1}$ highlighted with two black arrows, whereas for long 5-AGNRs (black) it shows a single peak at 1565 cm$^{-1}$. To further understand the origin of these peaks we used density functional theory (DFT) to investigate the vibrational modes including their length-dependent Raman features. The calculations for the periodic (infinitely long GNR) are based on a DFT implementation of the quantum mechanical treatment of resonant Raman employing a localized atomic orbital (LCAO) basis set[45,46]. The results in Figure 3b clearly show higher intensity for the LO mode in comparison to the TO mode, and therefore the former is more likely to be observed experimentally and thus corresponds to the single G peak for 5-AGNRs at 1565 cm$^{-1}$ in Figure 3a.

The full quantum mechanical treatment of Raman intensity is computationally not tractable for large-size unit cells such as those needed to model finite-size AGNRs. To this end, we use the bond-polarization model to compute the semi-classical non-resonant Raman signal[47] for finite-length GNRs. In Figure 3c we show the calculated LO Raman intensities for the periodic 5-AGNR, 5-AGNRs with 2 (dimer) and 3 (trimer) perylene units and the sum of the Raman intensities of the dimer and the trimer. It becomes clear from the simulations that the LO mode is length-dependent, especially for small lengths, and that it can be observed at different frequencies for perylene dimers (1531 cm$^{-1}$) and trimers (1516 cm$^{-1}$).

Although the evolution of this shift is not fully understood, the extremely short length of these structures could account for the fact that we do not see a monotonic convergence to the fully periodic situation. However, we observe a similar behavior experimentally, with two LO modes at 1563 cm$^{-1}$ and 1531 cm$^{-1}$ tentatively assigned to 5-AGNRs of length 1.7 nm (dimer) and 2.55 nm (trimer), respectively.

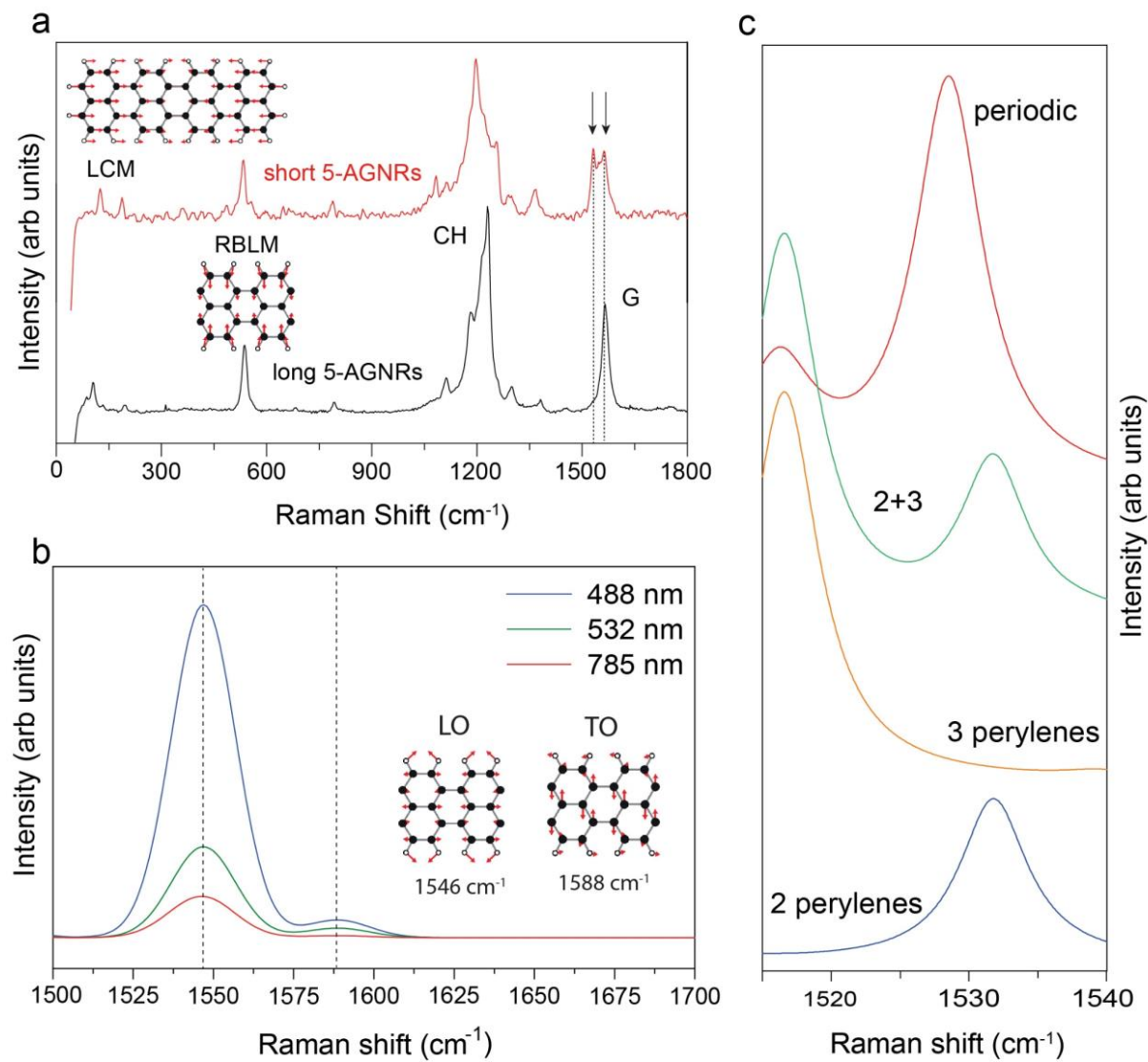

**Figure 3**. a) Experimental Raman spectra of short (red) and long (black) 5-AGNRs measured in vacuum (base pressure of $10^{-2}$ mbar) with 785 nm laser (maps of $10\times10$ µm$^2$, integration time 25 s. insets: normal mode analysis of RBLM and LCM modes. The two dashed lines highlight the appearance of two peaks in the G region for short 5-AGNRs. b) Resonant Raman simulation of the LO and TO intensities with 785, 532 and 488 nm using DFT with an LCAO basis; inset: normal mode analysis of LO and TO. c) Simulated LO frequencies as a function of 5-AGNR length using semi-classical (non-resonant) Raman and DFT with a plane-wave basis. The use of different basis set explains why the frequencies of the LO modes do not exactly match between panels (b) and (c).

**Device integration and electrical characterization**

The tunable bandgap[3] and the atomic precision achieved through the bottom-up fabrication approach make AGNRs highly appealing for their use as the active material in room temperature switching devices[21]. Besides their stability in ambient conditions and during substrate transfer steps, GNRs must be of sufficient length to allow for efficient bridging between the electrodes.

Field-effect transistors using high dielectric constant and metallic electrodes have shown very efficient gating of AGNRs, leading to high-performance GNR devices at room temperature[21]. Here we use the transistor structure presented in Figure 4a,b that consists of a local bottom gate geometry of ~8 nm thick W gate capped with a ~5.5 nm $HfO_2$ gate dielectric (dielectric constant = 25) and of ~12 nm thickness drain (D) and source (S) Pd electrodes (~30-200 nm wide, ~15-20 nm gaps). The devices are fabricated from two high coverage 5-AGNRs samples, one using uniaxially aligned 5-AGNRs achieved through growth on a Au(788) substrate with narrow (111) terraces (Figure S8a), and one using randomly oriented 5-AGNRs grown on an Au(111)/mica substrate (Figure 1e). The aligned and randomly oriented GNRs are transferred for further device processing using the electrochemical delamination method[17,43] and polymer-free method[15], respectively. The well-preserved fingerprint Raman peaks (Figure S8b) suggest the absence of significant structural modification for 5-AGNRs after substrate transfer.

Figure 4c represents typical transfer characteristics (drain current, $I_D$, versus gate voltage, $V_{GS}$) of the aligned and non-aligned 5-AGNR FET devices at room temperature under vacuum. Both devices exhibit p-type FET behavior, alike FETs made from bottom-up synthesized 9-AGNRs and 7-AGNRs using Pd electrodes[21,48]. While the off-state currents of both devices are at the same gate leakage level, the aligned device exhibits higher on-state performance resulting in an on/off current ratio exceeding $10^3$ that is the highest reported for bottom-up synthesized AGNRs of the low-bandgap 3p+2 family to date[17,22,49]. This confirms that long 5-AGNRs are not metallic and that previously observed metallic behavior in very short GNRs (< 5 nm)[17,18] is indeed due to an additional transport channel stemming from the overlap of in-gap states localized at the zigzag termini of the nanoribbons.

As shown by Lawrence et al the end states of 5-AGNRs undergo a transition from spin-split end states (8 perylene units and above) to closed-shell states because they hybridize for shorter ribbons (less than 7 perylene units)[18]. We confirm this finding by GW calculations[50], where we find a transition of the HOMO-LUMO gap from closed-shell to open-shell 5-AGNRs as a function of ribbon length (Figure S9). The transport gap for 5-AGNRs with 20 perylene units (17 nm) is around 1.7 V.

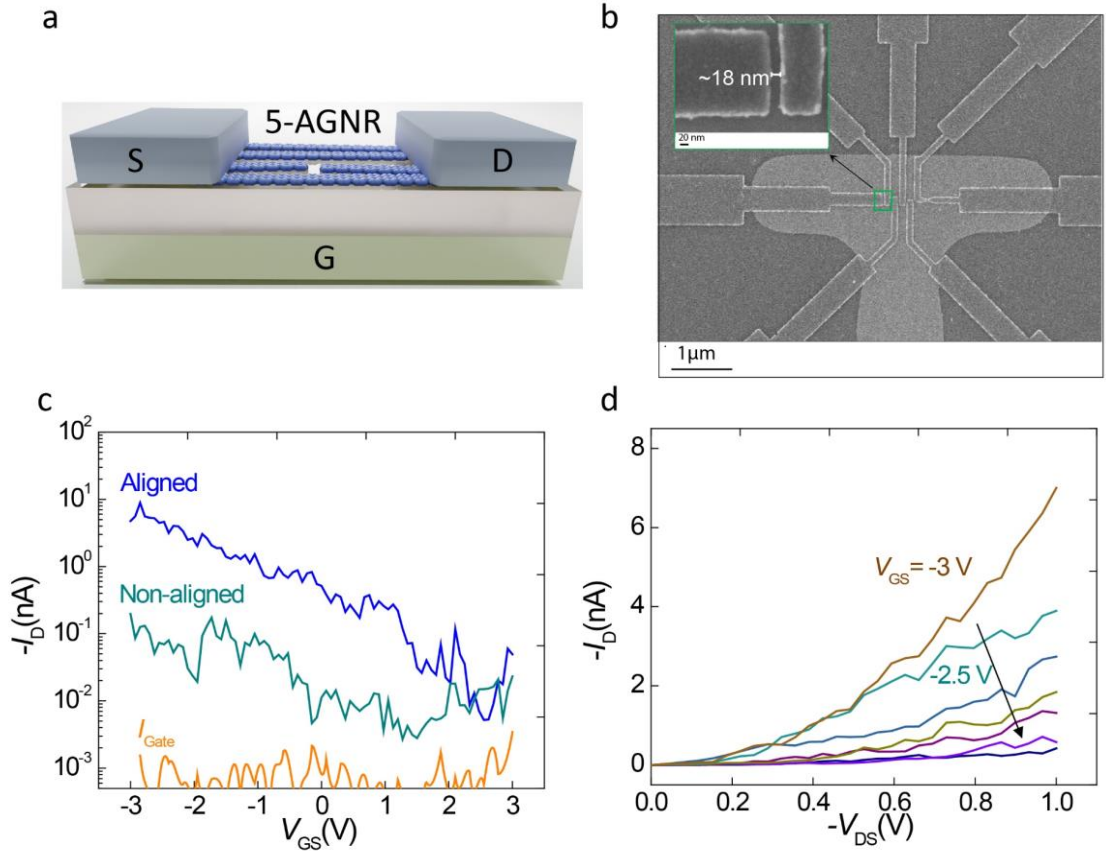

**Figure 4.** FETs made with bottom-up synthesized 5-AGNRs. **(a)** Schematic of 5-AGNR FET device. **(b)** SEM image of 5-AGNR devices fabricated on $HfO_2$ gate dielectric and W local bottom gate with Pd electrodes. The inset is a high magnification SEM image of the source-drain gap. **(c)** $I_D - V_{GS}$ characteristics of an aligned and non-aligned 5-AGNR FET at $V_{DS}$= -1 V at room temperature. **(d)** $I_D - V_{DS}$ characteristic of the same aligned device with varying $V_{GS}$.

The working device yield of the aligned 5-AGNR devices is also much higher than that of the non-aligned GNR devices, 82% vs ~12%, respectively. Both the on-current and device yield improvements can be explained by the larger number of GNRs bridging the channel in the aligned GNR configuration. Figure 4d shows the typical output characteristic ($I_D$ versus source voltage $V_{DS}$) of the aligned GNR device. The super-linearity of the $I_D$-$V_D$ indicates the presence of a finite Schottky Barrier (SB) at the Pd-GNR interface, which is limiting the performance of the 5-AGNR devices, as also commonly reported for FETs made with other bottom-up synthesized GNRs[21,48,51]. The work-function engineering of the contact metal or doping of the contact area could potentially eliminate the SB effects[52].

## Conclusions

In this work we used a combination of scanning tunneling microscopy and temperature-programmed X-ray photoelectron spectroscopy to unveil the reaction mechanisms leading to the formation of 5-AGNRs with different lengths, as identified with STM and by spectral differences in both LCM and G Raman modes. We achieved long 5-AGNRs because of the use of i) iodinated precursor molecules and ii) initial precursor coverage exceeding one monolayer, which lead to iii) the combined desorption of hydrogen released during the cyclodehydrogenation process with iodine and thus reduced hydrogen passivation of growing GNRs. Due to their increased length, the 5-AGNRs could be successfully integrated into FET devices with a fabrication yield of 82% and $I_{on}/I_{off}$ ratios exceeding $10^3$. Our results underline the importance of a microscopic understanding of on-surface reactions to achieve GNR layers with the desired properties to obtain switching devices that can be operated at room temperature.

## Materials and Methods

### Synthesis of dibromoperylene (DBP) and diiodoperylene (DIP).

DBP was synthesized according to the reported procedure[6], affording an isomeric mixture of 3,9-dibromoperylene and 3,10-dibromoperylene. By converting Bromo groups of DBP to Iodo groups, a mixture of 3,9-diiodoperylene and 3,10-diiodoperylene was obtained in 89% yield, which was used for the on-surface synthesis. The characterization data and experimental details are described in the SI.

### Scanning Tunneling Microscopy

A commercial low-temperature STM (Scienta Omicron) system was used for sample preparation and in situ characterization under ultra-high vacuum conditions. The Au(111) and Au(788) single crystals (MaTeck) were cleaned by several cycles of Ar$^+$ sputtering (1 keV) and annealing (470 ºC) until clean surfaces with monoatomic terrace steps were achieved. Deposition of both the DBP and DIP precursors were performed by thermal sublimation from a 6-fold organic evaporator. STM images were recorded in constant-current mode.

The molecular coverages of the samples in Figure 1 (with coverage less than 1 ML) are determined by inspecting the STM images of the sample after depositing the precursor molecules on the Au(111) surface. Therefore, the relation between the molecular sublimation time at a stable molecular flux (monitored by a quartz microbalance) and the molecular coverage can be well established. Then, the coverage of the samples, which is over 1 ML, can also be estimated from the molecular sublimation time. The molecular coverage of the samples in Figure 2 is extracted by examining the STM images of the samples and the areas of the XPS C 1s intensities with respect to the Au 4f intensities.

**X-ray photoelectron spectroscopy**

XPS measurements were performed at the X03DA beamline (PEARL endstation) at the SLS synchrotron radiation facility (Villigen, Switzerland), using linearly (and partially circularly left/right) polarized radiation with a photon energy of 425 eV. XPS spectra were obtained in normal emission geometry, using a hemispherical electron analyzer equipped with a multichannel plate (MCP) detector. HR-XPS spectra were recorded in "swept" mode with 20 eV pass energy. The TP-XPS measurement was performed during the heating of the sample (constant heating rate of 0.2 K·s$^{-1}$) using the "fixed" mode (snapshots of the Br 3d, I 4d and C 1s core levels) acquiring each spectrum for 5 s with 50 eV pass energy. The TP-XPS maps have a resolution of 3.5 °C in temperature and 17 s in time.

**Raman Spectroscopy**

Raman spectra were acquired in a homebuilt vacuum chamber with a WITec Alpha 300 R confocal Raman microscope in backscattering geometry. Measurements were performed with 785 nm laser excitation and a 300 g/mm grating on 5-GNRs on Au(111), Au(788) and SiO$_2$-based substrates. For maximum signal intensity, the laser power was maximized while avoiding optical damage[43] and spectra are recorded with a Zeiss 50x LD objective, NA = 0.55, through an uncoated fused silica window of only 0.2 mm thickness covering a hole of just 7 mm diameter. The vacuum chamber was mounted on a piezo stage for scanning.

**Device fabrication**

*Preparation of Local Bottom Gate Chips.* The local bottom gates were defined on a 100 nm SiO$_2$/Si wafer by sputtering ~8 nm W and subsequent patterning using photolithography and H$_2$O$_2$ wet-etch. The ~5.5 nm HfO$_2$ dielectric layer (Equivalent oxide thickness, EOT = ~1.8 nm, dielectric capacitance, C$_{ox}$ = ~ 19.0 µF cm$^{-2}$) was grown by atomic layer deposition (ALD) at 135 °C. Alignment markers and large pads for electrical probing were patterned using photolithography followed by lift-off of ~3 nm Cr and ~25 nm Pt. The wafer was then diced, and individual chips were used for further device fabrication.

*Patterning of Pd Electrodes.* After the GNR transfer[15,43], poly(methyl methacrylate) (molecular weight 950 kDa) was spun on the chips at 4500 rpm followed by a 10 min bake at 180 °C. Next, the drain and source electrodes (~30-200 nm wide, ~15-20 nm gaps) were patterned using a JEOL 6300-FS 100 kV EBL system and subsequently developed in 3:1 IPA-MIBK at 5 °C for 90 s. Finally, ~12 nm Pd was deposited using e-beam evaporation (Kurt J. Lesker) under ~$10^{-8}$ torr vacuum, and lift-off was completed in a Remover PG at 80 °C.

*Electrical Characterization*. The electrical characterization of the devices was performed in a Lakeshore TTPX cryogenic probe station with a vacuum level of <$10^{-5}$ torr at room temperature, using an Agilent B1500 semiconductor parameter analyzer.

**Acknowledgement**

We acknowledge funding by the Swiss National Science Foundation under grant no. 200020_182015 and 159690, the European Union Horizon 2020 research and innovation program under grant agreement no. 881603 (GrapheneFlagship Core 3), and the Office of Naval Research BRC Program under the grant N00014-18-1-2708. J.W and C. P. acknowledge the Gauss Centre for Supercomputing for providing computational resources on SuperMUC-NG under the project ID pn72pa. K.M. acknowledges a fellowship from Gutenberg Research College, Johannes Gutenberg University Mainz.

This work was also supported in part by the Office of Naval Research (ONR) MURI Program N00014-16-1-2921, the NSF Center for Energy Efficient Electronics Science, and the NSF under award DMR-1839098. Additional support was provided by the Berkeley Emerging Technology Research (BETR) Center. Device fabrication was mainly performed at the Stanford Nano Shared Facilities (SNSF) at Stanford University, supported by the NSF under award ECCS-1542152. Part of the device fabrication and electron microscopy imaging was performed at the Marvell Nanofabrication Laboratory at the University of California, Berkeley (UCB), and at the Molecular Foundry at Lawrence Berkeley National Laboratory (LBNL), supported by the Office of Science, Office of Basic Energy Sciences, of the U.S. Department of Energy (DOE) under contract no. DE-AC02-05CH11231. M.L.P. acknowledges funding by the Swiss National Science Foundation (SNSF) under the Spark project no. 196795.

The XPS experiments were performed on the X03DA (PEARL) beamline at the Swiss Light Source, Paul Scherrer Institut, Villigen, Switzerland. We thank the beamline manager Matthias Muntwiler (PSI) for his support during the experiments. We also acknowledge Lukas Rotach for the excellent technical support during the experiments.

# Supplementary Information

# Growth optimization and device integration of narrow-bandgap graphene nanoribbons

Gabriela Borin Barin[1]*, Qiang Sun[1]*[+], Marco Di Giovannantonio[1][++], Cheng-Zhuo Du[2], Xiao-Ye Wang[2], Juan Pablo Llinas[3], Zafer Mutlu[3], Yuxuan Lin[3], Jan Wilhelm[4], Jan Overbeck[1], Colin Daniels[5], Michael Lamparski[5], Hafeesudeen Sahabudeen[6], Mickael L. Perrin[1], José I. Urgel[1][+++], Shantanu Mishra[1][++++], Amogh Kinikar[1], Roland Widmer[1], Samuel Stolz[1][+++++], Max Bommert[1], Carlo Pignedoli[1], Xinliang Feng[6], Michel Calame[1], Klaus Müllen[7,8], Akimitsu Narita[7,9], Vincent Meunier[5], Jeffrey Bokor[3], Roman Fasel[1,10] and Pascal Ruffieux[1]

[1]Empa, Swiss Federal Laboratories for Materials Science and Technology, 8600 Dübendorf, Switzerland

[2]State Key Laboratory of Elemento-Organic Chemistry, College of Chemistry, Nankai University, Tianjin 300071, China

[3]Department of Electrical Engineering and Computer Sciences, University of California, Berkeley, CA 94720, United States

[4]Institute of Theoretical Physics, University of RegensburgD-93053 Regensburg, Germany

[5]Department of Physics, Applied Physics, and Astronomy, Rensselaer Polytechnic Institute, Troy, New York 12180, United States

[6]Center for Advancing Electronics Dresden, Department of Chemistry and Food Chemistry, TU Dresden, Dresden 01062, Germany

[7]Max Planck Institute for Polymer Research, 55128 Mainz, Germany

[8]Department of Chemistry, Johannes Gutenberg-Universität Mainz, 55128 Mainz, Germany

[9]Organic and Carbon Nanomaterials Unit, Okinawa Institute of Science and Technology Graduate University, 1919-1 Tancha, Onna-son, Okinawa 904-0495, Japan

[10]Department of Chemistry, Biochemistry and Pharmaceutical Sciences, University of Bern, 3012 Bern, Switzerland

[+]Present address: Materials Genome Institute, Shanghai University, Shanghai 200444, China

[++]Present address: Istituto di Struttura della Materia – CNR (ISM-CNR), via Fosso del Cavaliere 100, Roma 00133, Italy

[+++]Present address: IMDEA Nanoscience, C/Faraday 9, Campus de Cantoblanco, Madrid 28049, Spain

[++++]Present address: IBM Research – Zurich, Rüschlikon 8803, Switzerland

[+++++]Present address: Department of Physics, University of California, Berkeley, Berkeley, CA 94720, USA

**Synthesis of 3,9-diiodoperylene and 3,10-diiodoperylene (DIP).**

**General**

Unless otherwise noted, all chemicals were purchased from commercial sources and used as received without further purification. Nuclear magnetic resonance (NMR) spectra were recorded on Bruker AV 400 spectrometer. Chemical shifts ($\delta$) are reported in ppm. CDCl$_3$ or CD$_2$Cl$_2$ was used as the solvent and the internal chemical shift reference. High-resolution mass spectrometry (HRMS) was recorded on a Bruker solariX XR FT-ICR mass spectrometer with matrix-assisted laser desorption/ionization (MALDI) source and calibrated against 2,5-dihydroxybenzoic acid (DHB).

**Synthetic Procedure**

**Synthesis of 3,9-diiodoperylene and 3,10-diiodoperylene (DIP)**

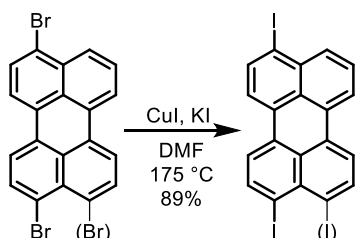

To a 100 mL two-necked round bottomed flask equipped with a reflux condenser was added DBP (0.41 g, 1.0 mmol), CuI (5.71 g, 30.0 mmol), KI (9.96 g, 60.0 mmol) and dry DMF (50 mL) under argon. The mixture was heated at 175°C for 90 h. After cooling down to room temperature, the reaction mixture was poured into water and filtered. The residue was dispersed in CH$_2$Cl$_2$ and then passed through Celite. After removal of the solvent under reduced pressure, the crude product was slowly recrystallized from hot toluene/ethanol to afford a mixture of 3,9-diiodoperylene and 3,10-diiodoperylene (DIP) as a yellow solid (448 mg, 89% yield). $^1$H NMR (400 MHz, CD$_2$Cl$_2$, 298 K, ppm) $\delta$ 8.32 – 8.27 (m, 2H), 8.13 – 8.10 (m, 2H), 8.02 – 7.99 (m, 2H), 7.94 – 7.90 (m, 2H), 7.63 – 7.57 (m, 2H); $^{13}$C NMR (101 MHz, CDCl$_3$, 298 K, ppm): $\delta$ 138.33, 138.25, 132.75, 132.55, 131.40, 128.22, 128.13, 121.78, 121.75, 121.51, 121.44. HRMS (MALDI-TOF) *m/z*: Calcd for C$_{20}$H$_{10}$I$_2$: 503.8872; Found: 503.8885.

**NMR Spectra**

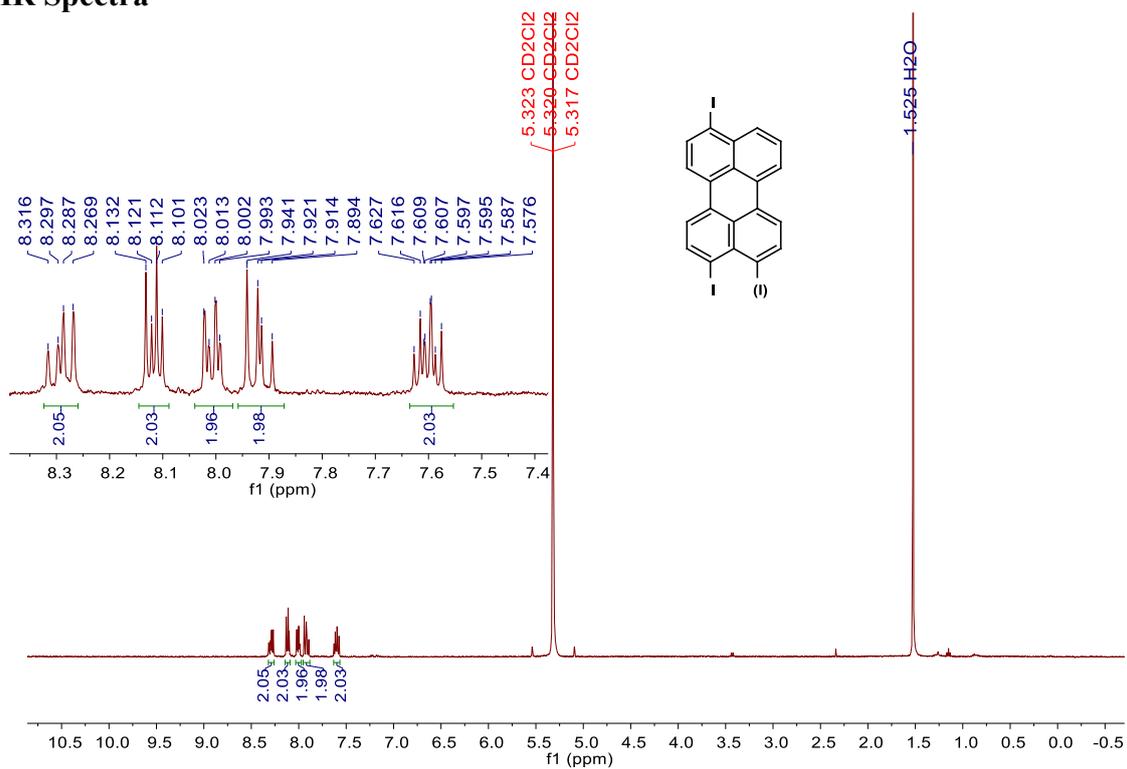

**Figure S1**. ¹H NMR spectrum of DIP (400 MHz, $CD_2Cl_2$, 298 K).

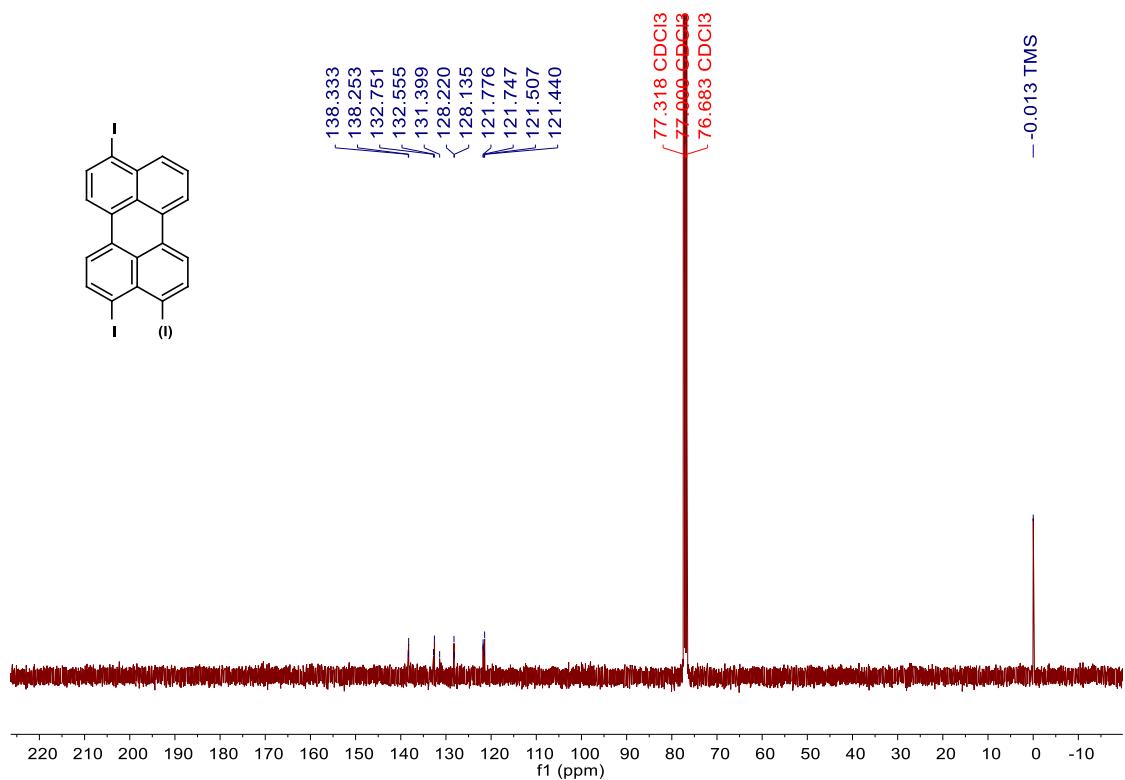

**Figure S2**. ¹³C NMR spectrum of DIP (101 MHz, $CDCl_3$, 298 K).

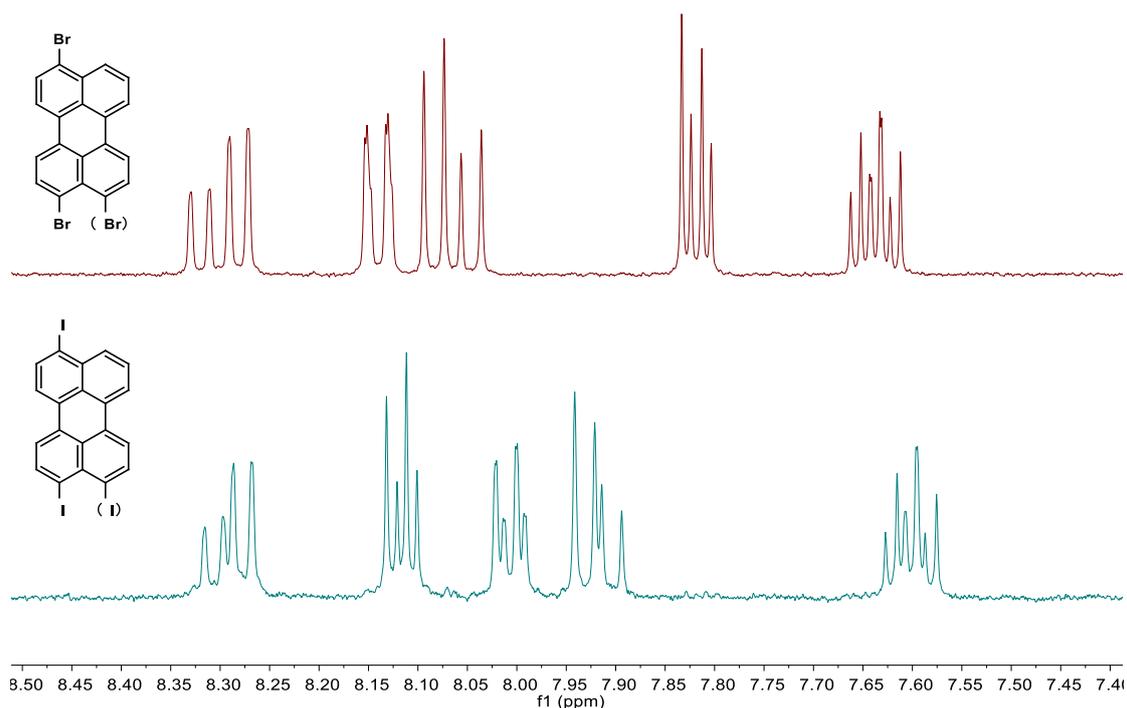

**Figure S3**. Stacked $^1$H NMR spectra (400 MHz, CD$_2$Cl$_2$, 298 K) of DBP and DIP.

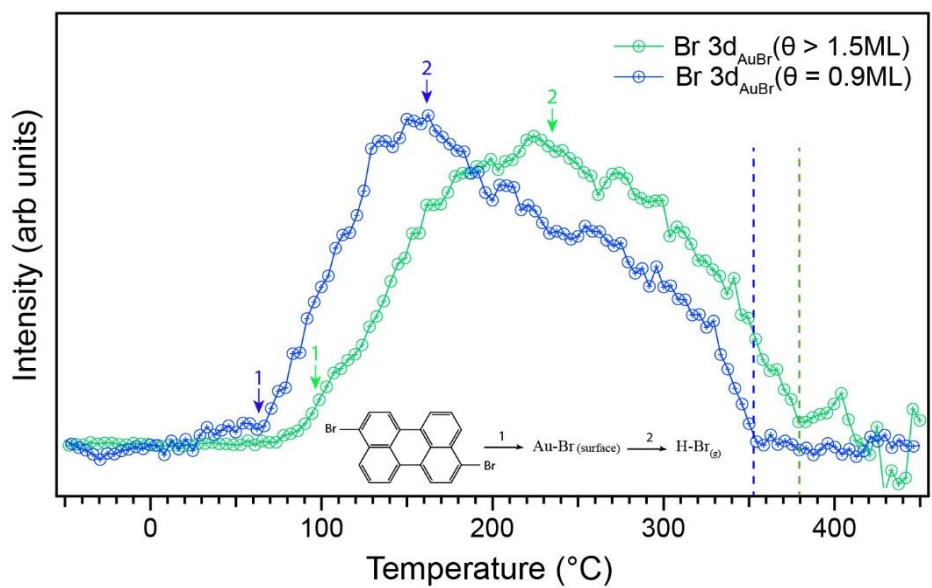

**Figure S4** Intensities of the Br 3d signal corresponding to the bromine atoms chemisorbed on Au(111) (Au-Br), as extracted from the Br 3d TP-XPS maps in Fig. 2a,b in the main text. The reaction scheme indicates different states of the brominated molecule during annealing, with the labels corresponding to those in the intensity curves. The dotted lines indicate complete bromine desorption from the surface.

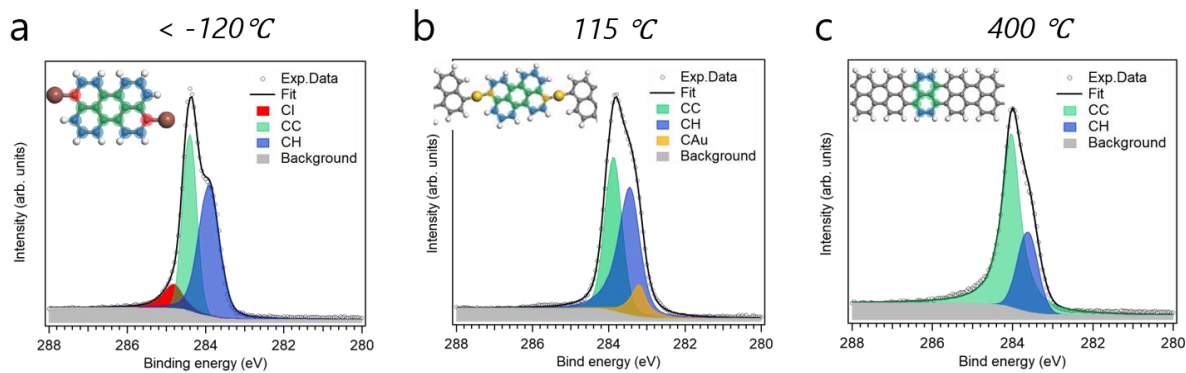

**Figure S5**. C 1s HR-XPS spectra of the sample corresponding to the TP-XPS map in Figure 2c in the main text. (a) Spectrum acquired after deposition of DIP onto Au(111) at a low temperature, which can be fit by the three expected carbon components in the intact molecular precursor, i.e., C-I, C-C, and C-H, with the expected C-I:C-C:C-H ratio of 1:4:5. (b) Spectrum acquired after annealing the surface to 115 °C. The spectrum can be fit by carbons in three different chemical environments (C-C, C-H and C-Au) with their relative areas corresponding to a C-Au-C organometallic chain, i.e., C-C:C-H:C-Au = 4:5:1. (c) Spectrum acquired after annealing the sample to 400 °C. The relative areas of the fitting components confirm the 5-AGNR, i.e., C-C:C-H = 3:2. The chemical models of three different phases are shown in the corresponding panels, and the different carbon species are highlighted in the model with colors that correspond to the fitting components.

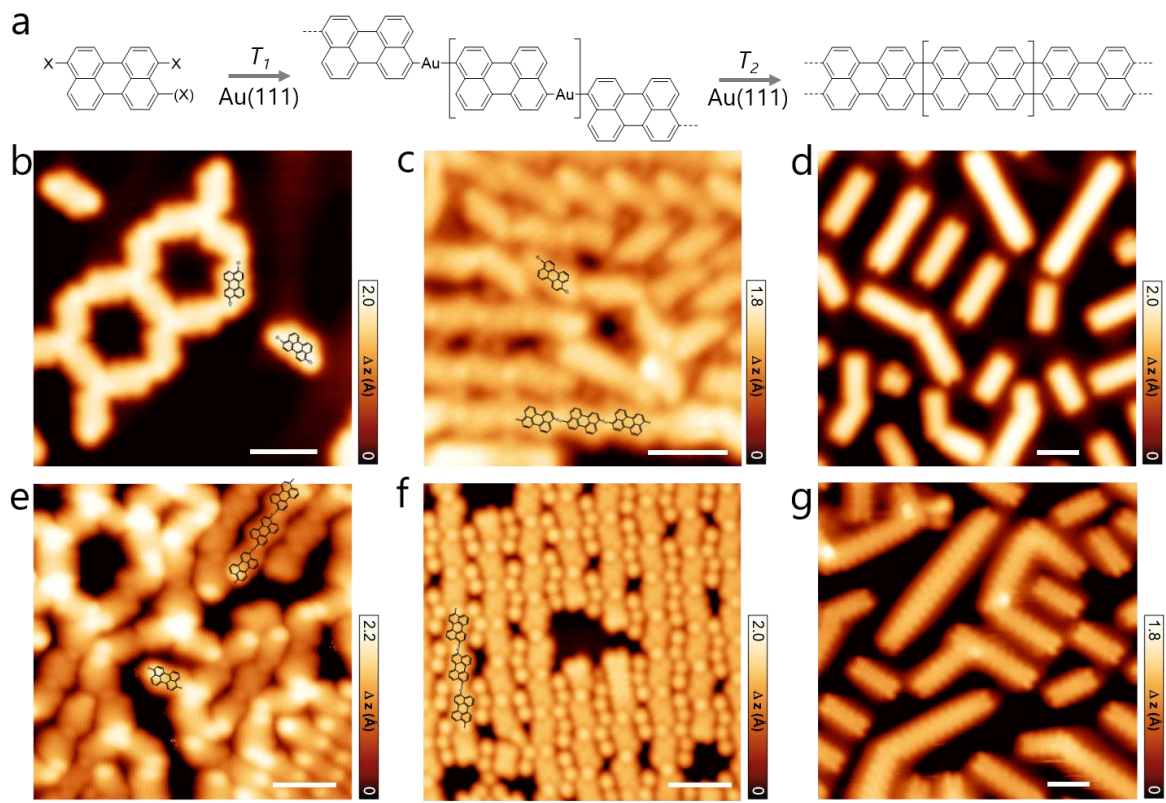

**Figure S6**. (a) Reaction scheme from the halogenated perylene precursors to 5-AGNRs, with an intermediate organometallic phase on Au(111). STM images after (b) depositing a submonolayer of DBP precursors on Au(111) at RT ($V_s$ = 1 V, $I_t$ = 100 pA), (c) annealing at 130 °C ($V_s$ = 15 mV, $I_t$ = 300 pA) and (d) 350 °C ($V_s$ = 0.1 V, $I_t$ = 200 pA). Intact molecules, short 5-AGNR segments and organometallic chains coexist in (c) due to the overlapping between the debromination and cyclodehydrogenation processes. The lower panels show STM images after (e) depositing a submonolayer of DIP precursors on Au(111) at RT ($V_s$ = -50 mV, $I_t$ = 150 pA), (f) annealing at 115 °C ($V_s$ = -0.1 V, $I_t$ = 200 pA) and (g) 350 °C ($V_s$ = -1.2 V, $I_t$ = 100 pA). Organometallic chains are already observed in the RT phase in (e). A pure phase of the organometallic chain are present after a mild annealing process in (f). Scale bars: 2 nm.

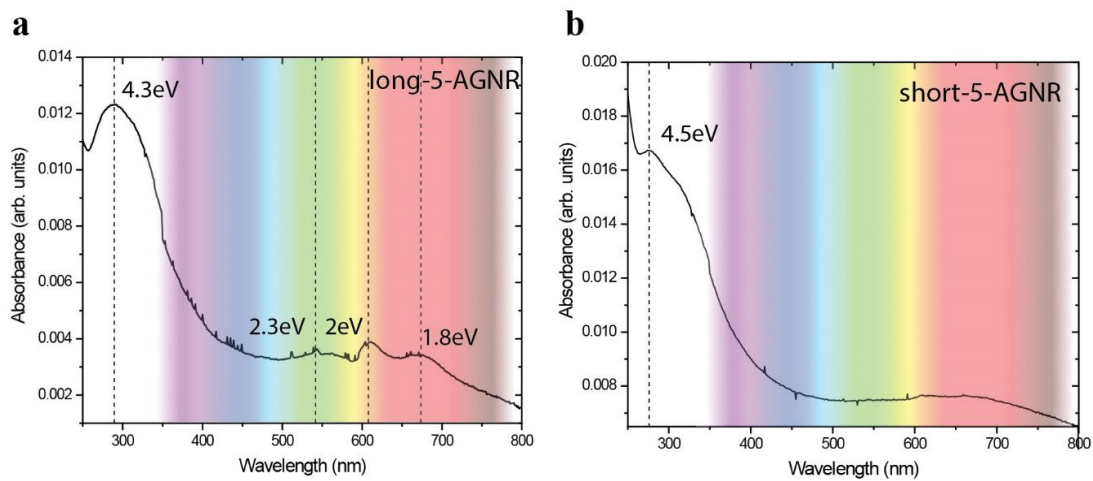

**Figure S7**. UV-vis spectroscopy of 4 layers of transferred GNRs on $Al_2O_3$: a) long 5-AGNRs and b) short 5-AGNRs.

The optical absorption of AGNRs can be tuned over a wide range by changing the GNR width[1–3]. This tunability opens up possibilities to use GNR as a platform in sensing and detecting from UV to IR applications. Here, we investigate the optical properties of 5-AGNRs via UV-vis spectroscopy of both short and long 5-AGNRs synthesized from DIP transferred onto $Al_2O_3$ substrates via a polymer-free transfer method, as previously reported[4,5]. For each sample, 4 layers of 5-AGNRs were transferred sequentially in order to improve the signal to noise ratio and their absorption spectra were measured in the UV−vis range (200−800 nm) in transmission geometry.

Long 5-AGNRs were demonstrated to have a bandgap of 0.85 eV on Au(111)[6]. In very short 5-AGNRs, on the other hand, the zig-zag termini induce end-states with energies that are located within the bulk band gap of 5-AGNRs and hence reduce energies between highest occupied and lowest unoccupied states to few hundreds of meV[6,7]. In a recent study of CVD grown 5-AGNR films Chen *et al* measured optical bandgaps in the range of 0.8-0.9 eV for long ribbons[1], showing a similar magnitude of the electronic band gap[8]. In our case, the absorbance profiles for short/long GNRs are clearly different, with clear absorption features observed for the long 5-AGNRs at 4.3, 2.3, 2.0 and 1.8 eV (288, 543, 608 and 671 nm, respectively). For the short 5-AGNRs we observe only a high optical transition at 4.5 eV (276 nm) but lower optical transitions are not visible.

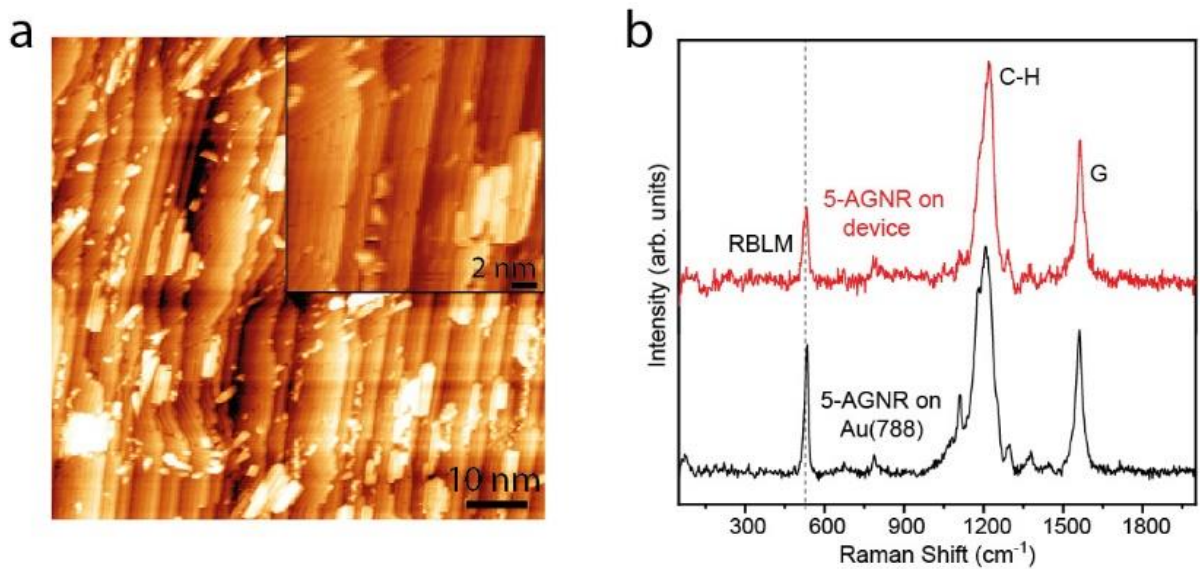

**Figure S8:** STM image of aligned 5-AGNR on Au(788) ($V_s$ = -1.5V, $I_t$ = 0.03 nA), inset showing a zoom-in of a region with aligned GNRs and GNRs on the second layer, b) Raman profile of 5-AGNRs on Au(788) and on device after substrate transfer

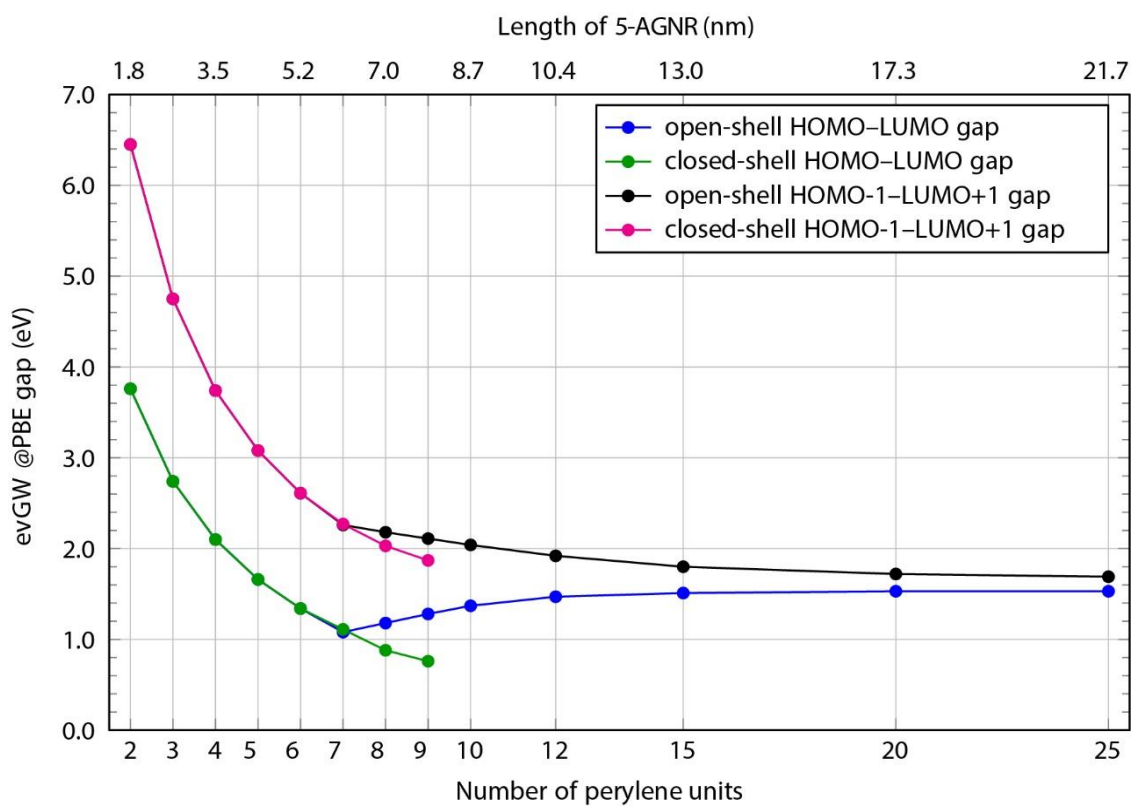

**Figure S9**. HOMO-LUMO and HOMO-1-LUMO+1 gaps computed from evGW@PBE for 5-AGNRs as function of the ribbon length starting from a closed-shell and open-shell PBE ground state.

For the eigenvalue-self consistent *GW* (ev*GW*) calculations, we have used the CP2K code[9–11]. We start from all-electron DFT calculations[12] using the Perdew-Burke-Ernzerhof (PBE) exchange correlation functional[13] and an augmented correlation-consistent double-zeta basis[14].

Two different electronic ground states from density functional theory have been employed in the GW calculations: First, a closed-shell (CS) ground state where the spatial orbitals for alpha (spin up) and beta (spin down) electrons are identical, and second, a broken-symmetry, singlet open-shell (SOS) ground state employing a single determinant where the spatial orbitals for alpha and beta electrons may differ. It is found that the SOS ground state and the CS ground state from DFT are identical for 5-AGNRs with up to 7 perylene units, which results in matching CS and SOS GW gaps up to this length. For 5-AGNRs consisting of 8 and 9 perylene units, the SOS ground state is found to be lower in energy than the CS ground state, with a larger GW gap for the SOS ground state. The increasing HOMO-LUMO gap from 7 to 15 perylene units may be an artifact of the spurious PBE starting point that employs a single Slater determinant instead of a superposition of Slater determinants.

The gap governing the transport properties of 5-AGNRs ("transport gap") is determined by the smallest gap between occupied and unoccupied molecular orbitals that are both delocalized over the whole 5-AGNR. For 5-AGNRs up to 7 perylene units, this is the HOMO-LUMO gap, while for 5-AGNRs with 8 or more perylene units it is the HOMO-1–LUMO+1 gap, since HOMO and LUMO are localized end states. The transport gaps of 5-AGNRs with 20 and 25 units is computed as 1.72 eV and 1.69 eV respectively, see Figure S9, such that we determine the transport gap to be 1.7 eV for 5-AGNRs with a length of 15 nm to 20 nm as present in the experiment.